\documentclass[showpacs,aps,graphicx,twocolumn]{revtex4}
\usepackage{graphicx}

\begin{document}

\title{ Efficient and economic five-party quantum state sharing of an arbitrary m-qubit state}
\author{ Yu-Bo Sheng,$^{1,2,3}$ Fu-Guo Deng,$^{1,2,3,4}$\footnote{Electronic mail: fgdeng@bnu.edu.cn}
 and Hong-Yu Zhou$^{1,2,3}$}
\address{$^1$ The Key Laboratory of Beam Technology and Material
Modification of Ministry of Education, Beijing Normal University,
Beijing 100875,  China\\
$^2$ Institute of Low Energy Nuclear Physics, and Department of
Material Science and Engineering, Beijing Normal University,
Beijing 100875,  China\\
$^3$ Beijing Radiation Center, Beijing 100875,  China\\
$^4$ Department of Physics, Applied Optics Beijing Area Major
Laboratory, Beijing Normal University, Beijing 100875, China}
\date{\today }

\begin{abstract}
We present an efficient and economic scheme for five-party quantum
state sharing of an arbitrary m-qubit state with $2m$ three-particle
Greenberger-Horne-Zeilinger (GHZ) states and three-particle
GHZ-state measurements. It is more convenient than  other schemes as
it only resorts to three-particle GHZ states and three-particle
joint measurement, not five-particle entanglements and five-particle
joint measurements. Moreover, this symmetric scheme is in principle
secure even though the number of the dishonest agents is more than
one. Its total efficiency approaches the maximal value.
\end{abstract}
 \maketitle

\section{introduction}
In a secret sharing, a boss, say Alice wants to send a secret
message $M_A$ to her two agents, say Bob and Charlie who are far
away from Alice. Alice suspects that one of the two agents may be
dishonest and the dishonest one will do harm to her benefit if he
can obtain the secret message independently. Unfortunately, Alice
does not know who the dishonest agent is.  Alice believes that the
honest agent can prevent the dishonest one from destroying her
benefit if they act in concert. In classical secret sharing
crypto-system \cite{Blakley}, Alice splits her secret message $M_A$
into two pieces $M_B$ and $M_C$, and then sends them to Bob and
Charlie, respectively. When Bob and Charlie cooperate, they can read
out the message $M_A=M_B \oplus M_C$; otherwise, none can obtain a
useful information about the secret message. As classical signals
can be copied fully and freely, it is in principle  impossible for
Alice to transmit her secret message directly to her agents with
only classical physics. An alternative is that Alice first creates a
private key with each of the agents, and then encrypts the secret
message with one-time pad crypto-system before she sends it to her
agents. At present, quantum key distribution (QKD)
\cite{RMP,CORE,BidQKD,LongLiu,ABC} provides a secure way for
generating a private key between two authorized parties. With some
private keys, the three parties can share the secret message $M_A$
securely. Quantum secure direct communication
\cite{two-step,QOTP,Wangc,QSDCnetwork} in principle supplies a
secure way for transmitting the messages $M_B$ and $M_C$ directly
with quantum memory.

Quantum secret sharing (QSS) is the generalization of classical
secret sharing \cite{Blakley} into quantum scenario. There are two
main goals in QSS. One is used to share a  private key. The other is
used to share a quantum information, i.e., an unknown state.  In
1999, Hillery, Bu\v{z}ek and Berthiaume (HBB) \cite{HBB99} proposed
an original QSS scheme for sharing a private key with entangled
three-particle Greenberger-Horne-Zeilinger (GHZ) states.
Subsequently, Karlsson, Koashi and Imoto \cite{KKI} presented a QSS
scheme for creating a private key among three parties with
two-particle entangled states. Xiao et al. \cite{longqss}
generalized the HBB QSS scheme to the case with $N$ agents and also
gave out two ways for improving the efficiency of qubits in the QSS
scheme \cite{HBB99}. Now, there are a great number of QSS schemes
for sharing a private key, including the schemes
\cite{Bandyopadhyay,Karimipour,cpyang,dengqsscpl,dengqss,zhoup,chenp,Gottesman}
with entangled quantum systems and those
\cite{dengsinglephoton,Zhang,improving,yanpra,dengsinglephoton2,zhoupcpl}
with single photons.  When QSS is used to share an unknown state, it
has to resort to quantum entanglement
\cite{cleve,Peng,dengQSTS,controledteleportation,QSTS2,dengQSTS2,QSTS3,Gordon,Cheung,manep,zhangscp,Lance,lixhcpl,zhoucontrolled}.
In HBB QSS scheme \cite{HBB99}, the authors presented a scheme for
controlled teleporation of an arbitrary qubit, in which the receiver
can recover an unknown state only when he cooperate with the
controllers. In 1999, Cleve, Gottesman and Lo \cite{cleve} proposed
a scheme for sharing a quantum secret with three-dimensional quantum
states. In 2004, Lance et al. named the branch of quantum secret
sharing for quantum information "quantum-state sharing" (QSTS)
\cite{Lance}. In essence, QSTS equals to controlled teleporation
\cite{dengQSTS,controledteleportation,QSTS2,lixhcpl}. In 2004, Li et
al.\cite{Peng} introduced a scheme for sharing an unknown single
qubit with a multipartite joint measurement (i.e., multipartite
GHZ-state measurement). In 2005, Deng et al. proposed a symmetric
scheme for controlled teleportation of an arbitrary two-particle
state with a GHZ-state quantum channel \cite{controledteleportation}
and a QSTS scheme for sharing an arbitrary two-particle state with a
Bell-state quantum channel \cite{dengQSTS}. In 2006, Li et al.
\cite{QSTS2} proposed an efficient symmetric multiparty quantum
state sharing scheme for an arbitrary $m$-qubit state with a
GHZ-state quantum channel. Also, they generalized this scheme to the
case for sharing an unknown d-dimensional quantum system
\cite{lixhcpl}. In 2006, Deng et al. \cite{dengQSTS2} proposed a
circular QSTS scheme for sharing an arbitrary two-qubit state with
two-photon entanglements and Bell-state measurements. Now the models
for sharing an unknown state with a non-maximally entangled quantum
channel are studied by some groups
\cite{QSTS3,Gordon,Cheung,manep,zhoucontrolled}.

Although there are some QSTS schemes for sharing a single qubit or
an $m$-qubit quantum system, they are either not economic or
insecure for the case with two dishonest agents. For instance, the
schemes in Refs.
\cite{HBB99,longqss,controledteleportation,dengQSTS,QSTS2,lixhcpl,Cheung,manep,QSTS3,zhoucontrolled}
require a five-particle GHZ-state quantum channel for sharing a
unknown single-qubit state when they are used for five-party
quantum state sharing, that in Ref. \cite{Peng} requires
five-particle GHZ-state measurements, and the schemes in Refs.
\cite{dengQSTS2,zhangscp} cannot prevent two dishonest agents from
eavesdropping the message freely when they cooperate. In this
paper, we will present an efficient and economic five-party QSTS
scheme for sharing an arbitrary $m$-qubit state. It only resorts
to a three-particle GHZ-state quantum channel and three-particle
GHZ-state measurements, not a five-particle GHZ-state quantum
channel \cite{controledteleportation} or five-particle GHZ-state
joint measurements \cite{Peng}. Moreover, this scheme is secure if
the number of the dishonest agents is more than one (no more than
three). Except for the sender Alice, all the agents need only to
take $m$ single-particle measurements on their particles for
controlling the receiver to reconstruct the unknown quantum state.
It is more convenient than others in a practical application. As
almost all the quantum resource can be used to sharing the quantum
information and the classical information exchanged is minimal,
the total efficiency in this scheme approaches the maximal value.

\section{Economic five-party QSTS scheme for sharing a single-qubit state}

For three-particle maximally entangled quantum systems, the eight
GHZ states can be written as follows:
\begin{eqnarray}
\vert \Psi_{0}\rangle_{ABC}=\frac{1}{\sqrt{2}}(\vert 000\rangle + \vert 111\rangle)_{ABC}, \\
\vert \Psi_{1}\rangle_{ABC}=\frac{1}{\sqrt{2}}(\vert 000\rangle - \vert 111\rangle)_{ABC}, \\
\vert \Psi_{2}\rangle_{ABC}=\frac{1}{\sqrt{2}}(\vert 001\rangle + \vert 110\rangle)_{ABC}, \\
\vert \Psi_{3}\rangle_{ABC}=\frac{1}{\sqrt{2}}(\vert 001\rangle - \vert 110\rangle)_{ABC}, \\
\vert \Psi_{4}\rangle_{ABC}=\frac{1}{\sqrt{2}}(\vert 010\rangle + \vert 101\rangle)_{ABC}, \\
\vert \Psi_{5}\rangle_{ABC}=\frac{1}{\sqrt{2}}(\vert 010\rangle - \vert 101\rangle)_{ABC}, \\
\vert \Psi_{6}\rangle_{ABC}=\frac{1}{\sqrt{2}}(\vert 011\rangle + \vert 100\rangle)_{ABC}, \\
\vert \Psi_{7}\rangle_{ABC}=\frac{1}{\sqrt{2}}(\vert 011\rangle -
\vert 100\rangle)_{ABC},
\end{eqnarray}
where $\vert 0\rangle$ and $\vert 1\rangle$ are the two
eigenstates of the Pauli operator $\sigma_z$, called it Z
measuring basis (MB) (for example, the polarization of photons
along the z-direction, and $\vert 0\rangle$ and $\vert 1\rangle$
represent the horizontal and the vertical polarizations).

For sharing an arbitrary qubit $x$ which is in the state $\vert
\chi\rangle_{x}=\alpha \vert 0\rangle + \beta \vert 1\rangle$ among
the five parties, say Alice, Bob$_i$ ($i=1,2,3$), and Charlie, the
boss Alice first shares two three-particle GHZ states $\vert
\Psi_{0}\rangle$ with her four agents (see Fig.1). That is, Alice
shares a three-particle GHZ state $\vert
\Psi_{0}\rangle_{A_1B_1B_2}$ with Bob$_1$ and Bob$_2$, and shares
another three-particle GHZ state $\vert \Psi_{0}\rangle_{A_2B_3C}$
with Bob$_3$ and Charlie. Alice keeps the particles $A_1$ and $A_2$.
Alice can share securely the GHZ states with her agents by using the
decoy-photon technique \cite{decoy}. In detail, when Alice wants to
share a sequence of three-particle GHZ states with her agents
Bob$_1$ and Bob$_2$ (or Bob$_3$ and Charlie), she prepares some
decoy photons which are randomly in one of the four states $\{\vert
0\rangle, \vert 1\rangle, \vert +x\rangle=\frac{1}{\sqrt{2}}(\vert
0\rangle +\vert 1\rangle), \vert -x\rangle=\frac{1}{\sqrt{2}}(\vert
0\rangle -\vert 1\rangle)\}$ and then inserts them in the two
sequences of the GHZ-state particles. Alice sends the two sequences
to her two agents, respectively. Alice and her agents can exploit
the decoy photons to check the security of the transmission
\cite{attack}.

\begin{figure}[!h]
\begin{center}
\includegraphics[width=8cm,angle=0]{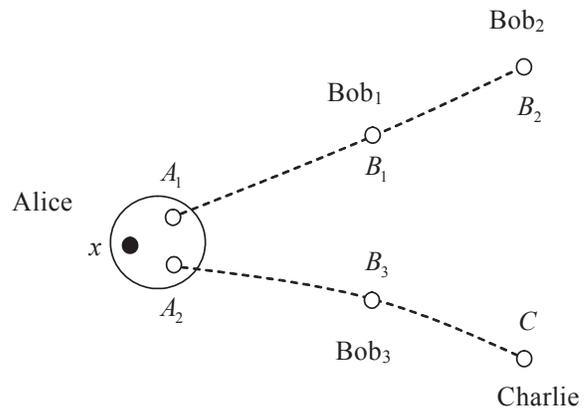}
\caption{ The principle of this QSTS scheme.  The three particles
linked with the dashed lines are in the GHZ state $\vert
\Psi_0\rangle$. The round represents a three-particle GHZ-state
measurement.}
\end{center}
\end{figure}

After setting up the quantum channel (two sequences of GHZ states)
with her agents, Alice can transfer her quantum information (the
unknown state) to the particles controlled by all the agents. In
detail, Alice performs a three-particle GHZ-state measurement on the
particles $x$, $A_1$, and $A_2$, and the quantum information of the
unknown qubit $x$ will be transferred into the subsystem composed of
the four particles $B_1$, $B_2$, $B_3$, and $C$. The four agents can
extract the quantum information with cooperation as the state of the
composite quantum system comprising the seven particles $x$, $A_1$,
$A_2$, $B_1$, $B_2$, $B_3$, and $C$ can be written as
\begin{widetext}
\begin{eqnarray}
\vert \Phi\rangle_s &\equiv& \vert \chi\rangle_{x}\otimes \vert
\Psi_{0}\rangle_{A_1B_1B_2} \otimes \vert
\Psi_{0}\rangle_{A_2B_3C}
\nonumber\\
&=&(\alpha \vert 0\rangle + \beta \vert 1\rangle)\otimes
\frac{1}{\sqrt{2}}(\vert 000\rangle + \vert 111\rangle)_{A_1B_1B_2}
\otimes \frac{1}{\sqrt{2}}(\vert 000\rangle + \vert
111\rangle)_{A_2B_3C}
 \nonumber\\
&=& \frac{1}{2\sqrt{2}}[\vert \Psi_0\rangle_{xA_1A_2}(\alpha \vert
00 \rangle_{B_1B_2} \vert 00 \rangle_{B_3C} + \beta \vert 11
\rangle_{B_1B_2} \vert 11 \rangle_{B_3C}) + \vert
\Psi_1\rangle_{xA_1A_2}(\alpha \vert 00 \rangle_{B_1B_2} \vert 00
\rangle_{B_3C} - \beta \vert 11 \rangle_{B_1B_2} \vert 11
\rangle_{B_3C})
\nonumber\\
&&\;\;\;\;\;\; +  \vert \Psi_2\rangle_{xA_1A_2}(\alpha \vert 00
\rangle_{B_1B_2} \vert 11 \rangle_{B_3C} + \beta \vert 11
\rangle_{B_1B_2} \vert 00 \rangle_{B_3C}) + \vert
\Psi_3\rangle_{xA_1A_2}(\alpha \vert 00\rangle_{B_1B_2} \vert
11\rangle_{B_3C} - \beta \vert 11\rangle_{B_1B_2} \vert 00
\rangle_{B_3C})
\nonumber\\
&&\;\;\;\;\;\; + \vert \Psi_4\rangle_{xA_1A_2}(\beta \vert
00\rangle_{B_1B_2} \vert 11\rangle_{B_3C} + \alpha \vert
11\rangle_{B_1B_2} \vert 00 \rangle_{B_3C}) - \vert
\Psi_5\rangle_{xA_1A_2}(\beta \vert 00\rangle_{B_1B_2} \vert
11\rangle_{B_3C} - \alpha \vert 11\rangle_{B_1B_2} \vert 00
\rangle_{B_3C})
\nonumber\\
&&\;\;\;\;\;\; + \vert \Psi_6\rangle_{xA_1A_2}(\beta \vert
00\rangle_{B_1B_2} \vert 00\rangle_{B_3C} + \alpha \vert
11\rangle_{B_1B_2} \vert 11 \rangle_{B_3C}) - \vert
\Psi_7\rangle_{xA_1A_2}(\beta \vert 00\rangle_{B_1B_2} \vert
00\rangle_{B_3C} - \alpha \vert 11\rangle_{B_1B_2} \vert 11
\rangle_{B_3C})].\nonumber\\
\end{eqnarray}
\end{widetext}
When Alice gets the outcome $\vert \Psi_0\rangle_{xA_1A_2}$, the
subsystem composed of the particles controlled by all the four
agents collapses to the state $\phi_0=\alpha \vert 00
\rangle_{B_1B_2} \vert 00 \rangle_{B_3C} + \beta \vert 11
\rangle_{B_1B_2} \vert 11 \rangle_{B_3C}$. The three controllers
Bob$_1$, Bob$_2$, and Bob$_3$ take a measurement with the basis $X$
on their particles $B_1$, $B_2$, and $B_3$, respectively. If the
number of the controllers who obtain the outcome $\vert +x\rangle$
is even, the particle $C$ will collapse to the state $\alpha \vert
0\rangle + \beta \vert 1\rangle$ and Charlie needs doing nothing on
his particle for recovering the originally unknown state $\vert
\chi\rangle$; otherwise, the state of the particle $C$ becomes
$\alpha \vert 0\rangle - \beta \vert 1\rangle$ and Charlie needs
performing a phase-flip operation $\sigma_z = \vert 0\rangle\langle
0\vert - \vert 1\rangle\langle 1\vert$ on the particle $C$ to
recover the unknown state $\vert \chi\rangle$.
\begin{widetext}
\begin{eqnarray}
\phi_0 &=& \alpha \vert 00 \rangle_{B_1B_2} \vert 00 \rangle_{B_3C}
+ \beta \vert 11 \rangle_{B_1B_2} \vert 11 \rangle_{B_3C}\nonumber\\
&=& \frac{1}{2\sqrt{2}}[(\vert +x\rangle\vert +x\rangle\vert
+x\rangle + \vert +x\rangle\vert -x\rangle\vert -x\rangle + \vert
-x\rangle\vert +x\rangle\vert -x\rangle + \vert -x\rangle\vert
-x\rangle\vert +x\rangle)_{B_1B_2B_3}(\alpha \vert 0\rangle + \beta
\vert
1\rangle)_C \nonumber \\
&&\;\;\;\;\;\;  + (\vert +x\rangle\vert +x\rangle\vert -x\rangle +
\vert +x\rangle\vert -x\rangle\vert +x\rangle + \vert -x\rangle\vert
+x\rangle\vert +x\rangle + \vert -x\rangle\vert -x\rangle\vert
-x\rangle)_{B_1B_2B_3}(\alpha \vert 0\rangle - \beta \vert
1\rangle)_C].\nonumber \\
\end{eqnarray}
\end{widetext}

\begin{center}
\begin{table}[!h]
 \caption{The relation between the unitary operations
used for recovering the unknown state and the outcomes obtained by
Alice, Bob$_1$, Bob$_2$, and Bob$_3$.}
\begin{tabular}{ccccc|cccc}\hline
$V_{xA_1A_2}$  & & $P_{total}$ & & & & $\phi_{C}$ & &
operations\\\hline
 0  & &  $+$ & &  & & $\alpha \vert
0\rangle + \beta\vert 1\rangle$ & & $I$ \\
 0  & &  $-$ & &  & & $\alpha \vert
0\rangle - \beta\vert 1\rangle$ & & $\sigma_z$ \\
 1  & &  $+$ & &  & & $\beta \vert
0\rangle + \alpha\vert 1\rangle$ & & $\sigma_x$ \\
 1  & &  $-$ & &  & & $ \beta\vert
0\rangle - \alpha\vert 1\rangle$ & & $i\sigma_y$
\\\hline
\end{tabular}\label{table1}
\end{table}
\end{center}

When Alice gets the other outcomes with GHZ-state measurements on
the particles $x$, $A_1$, and $A_2$, the relation between the
outcomes obtained by the controllers and the unitary operations
needed for recovering the unknown state $\vert \chi\rangle=\alpha
\vert 0\rangle + \beta \vert 1\rangle$ is shown in Table
\ref{table1}. Here $V_{xA_1A_2}$ is the value of the outcome
obtained by Alice. We code the states $\{\vert
\Psi_0\rangle_{xA_1A_2}, \vert \Psi_1\rangle_{xA_1A_2}, \vert
\Psi_4\rangle_{xA_1A_2}, \vert \Psi_5\rangle_{xA_1A_2}\}$ as 0 and
the states $\{\vert \Psi_2\rangle_{xA_1A_2}, \vert
\Psi_3\rangle_{xA_1A_2}, \vert \Psi_6\rangle_{xA_1A_2}, \vert
\Psi_7\rangle_{xA_1A_2}\}$ as 1. For example, $V_{xA_1A_2}=0$ if
Alice gets the outcome $\vert \Psi_0\rangle_{xA_1A_2}$ with her
GHZ-state measurement. In this table,
$P_{total}=P_AP_{B_1}P_{B_2}P_{B_3}$. Here $P_A$, $P_{B_1}$,
$P_{B_2}$, and $P_{B_3}$ are the parities of the outcomes obtained
by Alice, Bob$_1$, Bob$_2$, and Bob$_3$, respectively. Similar to
Refs. \cite{controledteleportation,dengQSTS,QSTS2}, we code the
parities of the states $\{\vert \Psi_0\rangle_{xA_1A_2}, \vert
\Psi_2\rangle_{xA_1A_2}, \vert \Psi_4\rangle_{xA_1A_2}, \vert
\Psi_6\rangle_{xA_1A_2}\}$ as $+$ and $\{\vert
\Psi_1\rangle_{xA_1A_2}, \vert \Psi_3\rangle_{xA_1A_2}, \vert
\Psi_5\rangle_{xA_1A_2}, \vert \Psi_7\rangle_{xA_1A_2}\}$ as $-$.
For the outcomes obtained by the controllers Bob$_1$, Bob$_2$, and
Bob$_3$, the state $\vert +x\rangle$ represents the parity $+$ and
the state $\vert -x\rangle$ represents the parity $-$. $\phi_C$ is
the state of the particle $C$ controlled by Charlie before the
unitary operation is done. $\sigma_z$, $\sigma_x$, and
 $\sigma_y$ are the Pauli operations, i.e.,
\begin{eqnarray}
I         &=&     \vert 0\rangle\langle 0\vert + \vert 1\rangle\langle 1\vert,\\
\sigma_z  &=&     \vert 0\rangle\langle 0\vert - \vert 1\rangle\langle 1\vert,\\
\sigma_x  &=&     \vert 0\rangle\langle 1\vert + \vert 1\rangle\langle 0\vert,\\
i\sigma_y &=&     \vert 0\rangle\langle 1\vert - \vert
1\rangle\langle 0\vert.
\end{eqnarray}

From Table \ref{table1}, one can see that Alice need only publish
two bits of classical information about her three-particle GHZ-state
measurement for her agents to recover the unknown state, not three
bits of classical information. Each of the controllers should
announce one bit of classical information about the outcome of the
measurement with the  basis $X$, and the receiver Charlie can
recover the unknown state $\vert \chi\rangle$ with a unitary
operation.

\section{Economic five-party QSTS scheme for sharing an arbitrary $m$-qubit state}

It is straightforward to generalize this five-party QSTS scheme to
the case for sharing an arbitrary $m$-qubit state. Same as Ref.
\cite{QSTS2}, an $m$-qubit state can be described as
\begin{eqnarray}
\vert \xi\rangle_{u}=\sum_{ij\ldots k}a_{\underbrace{ij\ldots
k}_m}\vert \underbrace{ij\ldots k}_m\rangle_{x_1x_2\ldots
x_m},\label{unknownstate2}
\end{eqnarray}
where $i, j, \ldots, k \in \{0,1\}$, and $x_1, x_2, \ldots$, and $
x_m$ are the $m$ particles in the unknown state. For sharing
$m$-qubit state $\vert \xi\rangle_{u}$, Alice should share at least
$m$ three-particle GHZ states $\vert \Psi_{0}\rangle$ with each two
of her agents, i.e., set up a quantum channel with $2m$ GHZ states
securely. The state of the composite quantum system composed of the
particles in the unknown $m$-qubit state and the GHZ states can be
written as
\begin{widetext}
\begin{eqnarray}
\left\vert\Psi\right\rangle&\equiv& (\sum_{ij\ldots
k}a_{\underbrace{ij\ldots k}_m}\vert \underbrace{ij\ldots
k}_m\rangle_{x_1x_2\ldots x_m}) \otimes
 \prod_{i'=1}^{m}[\frac{1}{2}(\left\vert 000\right\rangle + \left\vert 11
1\right\rangle)_{A_{1i'}B_{1i'}B_{2i'}} \otimes (\left\vert
000\right\rangle + \left\vert 11
1\right\rangle)_{A_{2i'}B_{3i'}C_{i'}}].
\end{eqnarray}
\end{widetext}
Alice can transfer the information of her unknown state into the
particles controlled by her four agents by performing $m$ GHZ-state
measurements on her particles. That is, she takes a GHZ-state
measurement on the particles $x_i$, $A_{1i}$, and $A_{2i}$, where
$i$ is the $i$-th particle in the unknown state or the $i$-th GHZ
state shared with her agents. Three agents can act as the
controllers and the other one acts as the receiver who can recover
the unknown $m$-qubit state with the help of all the controllers. In
this scheme, each of the controllers takes $m$ single-particle
measurements on his particles with the basis $X$, and tells the
receiver his outcomes when they cooperate to recover the unknown
state $\vert \xi\rangle_u$.

\begin{table}[!h]
\caption{The relation between the values of $V_i, P_i$ and the local
unitary operations $U_i$. }
\begin{tabular}{ccccccccc}\hline
$V_i$  &  & 0 &  & 0 & & 1 & & 1 \\
$P_i$  & &  $+$ & & $-$ & & $+$ & & $-$ \\
$U_i$ & &  $I$  & &  $\sigma_z$   & &  $\sigma_x$  & &  $i\sigma_y$\\
\hline
\end{tabular}\label{table2}
\end{table}

The relation between the outcomes of measurements and the local
unitary operations with which the receiver can recover the unknown
state is shown in Table \ref{table2}. That is, Charlie can
reconstruct the unknown state $\vert \xi\rangle$ according to the
Table \ref{table2} if he cooperates with all the controllers. Here
$V_i$ is the value of the outcomes of the $i$-th GHZ-state
measurement done by Alice. That is, $V_i=0$ if Alice takes a
GHZ-state measurement on the particles $x_i$, $A_{1i}$, and
$A_{2i}$, and obtains the outcomes $\{\vert
\Psi_0\rangle_{x_iA_{1i}A_{2i}}, \vert
\Psi_1\rangle_{x_iA_{1i}A_{2i}}, \vert
\Psi_4\rangle_{x_iA_{1i}A_{2i}}, \vert
\Psi_5\rangle_{x_iA_{1i}A_{2i}}\}$; otherwise, $V_i=1$. $P_i$ is the
product of the parities of all the outcomes in the $i$-th
measurements done by Alice and her agents, i.e.,
$P_i=P_{A_i}P_{B_{1i}}P_{B_{2i}}P_{B_{3i}}$.

With the information published by Alice and the three controllers,
the receiver, say Charlie, can recover the unknown state. In this
time, Charlie need only take the unitary operation $U_i$ on the
$i$-th ($i=1,2,\cdots, m$) particle controlled by him. After $m$
operations are performed on all his particles, Charlie obtains the
unknown state $\vert \xi\rangle_{u}$. Same as the case for sharing a
single-qubit state, this scheme for sharing $m$ qubits is secure if
the quantum channel, two sequences of three-particle GHZ states, is
set up securely.

\section{discussion and summary}

As three-particle GHZ states are maximally entangled ones, the
receiver Charlie can reconstruct the unknown $m$-qubit state $\vert
\xi\rangle_{u}$ with the probability 100\% in principle if he
cooperates with all the other agents, same as Ref. \cite{QSTS2}.
Certainly, without the outcomes obtained by the three controllers,
Charlie cannot recover the unknown state  $\vert \xi\rangle_{u}$
even though he obtains the outcome published by Alice. On the one
hand, Charlie does not know whether the controllers measure their
particles with the basis $X$ or not. That is, Charlie does not know
whether his particles $C_i$ still entangles with those controlled by
the three controllers or not. On the other hand, Charlie does not
know how to choose his unitary operations for recovering the unknown
state even though he knows the fact that all the controllers have
measured their particles but not the outcomes. That is, he will only
has the probability $\frac{1}{2^m}$ to get the correct result if one
of the three controllers does not agree to cooperate as Charlie has
only half of the chance to choose the correct operation for each
qubit $C_i$ according to the information published by Alice and the
other two controller, shown in Table \ref{table2}. In detail, when
Alice obtains the value $V_i=0$, Charlie should choose one of the
two unitary operations $\{I, \sigma_z\}$; otherwise, he should
choose one of the other two unitary operations $\{\sigma_x,
i\sigma_y\}$. Charlie can divide the four unitary operations into
two groups according to the value $V_i$, but he cannot determine
which one of two operations.  In a word, without the help of the
controllers, Charlie cannot reconstruct the originally unknown state
$\vert \xi\rangle_{u}$. That is, the security of this QSTS scheme is
the same as that of the quantum channel. As the quantum channel can
be set up with decoy-photon technique \cite{decoy} and multipartite
entanglement purification \cite{mep}, this QSTS is in principle
secure. Also, Alice can exploit the faithful-qubit-transmission
technique \cite{lixhapl} to improve the efficiency for setting up
the quantum channel in the condition with a collective noise.

In this five-party QSTS scheme, all the quantum sources (two
sequence of GHZ states shared) can be used to carry the quantum
information if all the agents act in concert after the quantum
channel is set up securely with the decoy-photon technique
\cite{decoy} and the faithful-qubit-transmission technique
\cite{lixhapl}. The proportion of the decoy photons is small and can
be neglectable in theory. That is, the intrinsic efficiency for
qubits $\eta_q \equiv \frac{q_u}{q_t}$ in this QSTS scheme
approaches 100\%, same as all other QSTS schemes based on maximally
entangled quantum channel
\cite{Peng,controledteleportation,dengQSTS,QSTS2,lixhcpl}. Here
$q_u$ is the number of the useful qubits in QSTS and $q_t$ is the
number of qubits transmitted. The total efficiency $\eta_t$ of QSTS
schemes can be calculated as follows \cite{QSTS2},
\begin{eqnarray}
\eta_t=\frac{q_u}{q_t+b_t},
\end{eqnarray}
where $b_t$ is the number of the classical bits exchanged for
sharing the unknown states. In this five-party QSTS scheme,
$q_u=q_t=4m$, and $b_t=5m$ as Alice announces $2m$ bits of outcomes
of the three-particle GHZ-state measurements and each of the three
controllers tells the receiver $m$ bits of outcomes of the
measurements with the basis $X$. That is, $\eta_t=\frac{4}{9}$ which
is the maximal value for QSTS \cite{QSTS2}, higher than that
($\frac{1}{3}$) in the QSTS scheme based on Bell states \cite{Peng}
in the case with four agents.

This five-party QSTS scheme for sharing an $m$-qubit state has some
advantages. First, it only resorts to three-particle GHZ-state
quantum channel, not a five-particle one as those in Refs.
\cite{HBB99,longqss,controledteleportation,dengQSTS,QSTS2,lixhcpl,Gordon,Cheung,manep,QSTS3,zhoucontrolled}.
In practical, it is more convenient than some other QSTS schemes as
it is more difficult for people to prepare five-particle
entanglements than three-particle entanglements
\cite{Mentanglement1,Mentanglement2,Mentanglement3}. Secondly, the
sender Alice need only perform three-particle GHZ-state measurements
on her particles, not five-particle GHZ-state joint measurements as
that in Ref. \cite{Peng}. Thirdly, this QSTS scheme is in principle
secure if the number of the dishonest agents is large than one (less
than four). That is, it does not require that at most one of the
agents is dishonest \cite{dengQSTS2,zhangscp}. Fourthly, the
controllers need only take $m$ single-particle measurements on their
particles for completing the task of controlling. Moreover, this
QSTS scheme is a symmetric one in which each of the agents can act
as the receiver who can recover the unknown state with the help of
the others. The amount of classical information exchanged in this
scheme is less than others, and its total efficiency approaches the
maximal value in theory.

Certainly, the QTS scheme shown in Ref. \cite{Peng} uses Bell
states as quantum channel for sharing a quantum information. If it
is used for sharing an unknown state with four agents, the sender
Alice should take a six-particle joint GHZ-state measurement on
her particle. At the aspect of resource, the scheme \cite{Peng} is
simper than the present one. However, it requires six-particle
joint GHZ-state measurements, which makes more difficult to be
implemented at experiment than the present one. Also the second
QSTS scheme in Ref.\cite{Gordon} exploits non-maximally
two-particle entangled states as quantum channel, which makes it
more convenient than the present one at the aspect of resource,
similar to that in Ref.\cite{Peng}. When it is used by the sender
Alice to share a unknown state with her four agents, six-particle
generalized GHZ-state measurements are required. At present, it is
very difficult to prepare an  entangled quantum system composed of
more than four particles
\cite{Mentanglement1,Mentanglement2,Mentanglement3}. On the other
hand, six-particle joint measurements are beyond what are
available at experiment at present. With development of technique,
those difficulties may be not the obstruct for implementing
multiparty quantum state sharing efficiently.

In summary, we have presented an efficient and economic scheme for
five parties to share an arbitrary $m$-qubit state with
three-particle GHZ states, not five-particle ones. The sender Alice
need only perform $m$ three-particle joint measurements on her
particles, not five-particle joint measurements, and each of the
three controllers need only take $m$ single-particle measurements on
his particles with the basis $X$. These two factors make this QSTS
scheme more convenient than others. As almost all the quantum
resource can be used to share the quantum information and the
classical information exchanged is minimal, the total efficiency in
this scheme approaches the maximal value $\frac{4}{9}$. Moreover,
this scheme is in principle secure even though there are more than
one dishonest agents (less than four), different from that with
two-photon entanglements and Bell-state measurements in Ref.
\cite{dengQSTS2}.

\bigskip

This work is supported by the National Natural Science Foundation
of China under Grant No. 10604008, A Foundation for the Author of
National Excellent Doctoral Dissertation of China under Grant No.
200723, and Beijing Education Committee under Grant No.
XK100270454.

\end{document}